\documentclass[twocolumn]{aastex62}

\usepackage{graphicx}				% Use pdf, png, jpg, or eps§ with pdflatex; use eps in DVI mode
\usepackage{xcolor}
\usepackage[sort&compress]{natbib}
\usepackage[hang,flushmargin]{footmisc}
\usepackage[counterclockwise]{rotating}

\newcommand{\teff}{$T_{\rm eff}$}
\newcommand{\logg}{$\log g$}
\newcommand{\feh}{$\mathrm{[Fe/H]}$}
\newcommand{\vt}{$v_t$}

\newcommand{\tc}{$T_\mathrm{C}$}

\newcommand{\dexK}{dex K$^{-1}$}

\newcommand{\I}{\textsc{I}}
\newcommand{\II}{\textsc{II}}
\newcommand{\acronym}[1]{{\small{#1}}}
\newcommand{\hip}[1]{{\acronym{HIP}\ #1}}

\shortauthors{Bedell et al.}
\shorttitle{Chemical Homogeneity of Sun-like Stars}

\begin{document}
\graphicspath{ {../figures/} }
\DeclareGraphicsExtensions{.pdf,.eps,.png}

%@arxiver{avgtwin_w_ncapture.pdf,tc_histogram_w_ncapture.pdf}

\title{\textsc{The Chemical Homogeneity of Sun-like Stars in the Solar Neighborhood}}

\author{Megan Bedell}
\affiliation{Center for Computational Astrophysics, Flatiron Institute, 162 5th Ave., New York, NY 10010, USA}
\affiliation{Department of Astronomy and Astrophysics, University of Chicago, 5640 S. Ellis Ave, Chicago, IL 60637, USA}

\author{Jacob L. Bean}
\affiliation{Department of Astronomy and Astrophysics, University of Chicago, 5640 S. Ellis Ave, Chicago, IL 60637, USA}

\author{Jorge Mel\'{e}ndez}
\affiliation{Departamento de Astronomia do IAG/USP, Universidade de
S\~{a}o Paulo, Rua do Mat\~{a}o 1226, Cidade Universit\'{a}ria, 05508-900 S\~{a}o Paulo, SP, Brazil}

\author{Lorenzo Spina}
\affiliation{Monash Centre for Astrophysics, School of Physics and Astronomy, Monash University, VIC 3800, Australia}
\affiliation{Departamento de Astronomia do IAG/USP, Universidade de
S\~{a}o Paulo, Rua do Mat\~{a}o 1226, Cidade Universit\'{a}ria, 05508-900 S\~{a}o Paulo, SP, Brazil}

\author{Ivan Ram\'{i}rez}
\affiliation{Tacoma Community College, 6501 South 19th Street, Tacoma, Washington 98466, USA}

\author{Martin Asplund}
\affiliation{Research School of Astronomy and Astrophysics, The Australian National University, Cotter Road, Canberra, ACT 2611, Australia}

\author{Alan Alves-Brito}
\affiliation{Universidade Federal do Rio Grande do Sul, Instituto de F\'{i}sica, Av. Bento Gon\c{c}alves 9500,  Porto Alegre, RS, Brazil}

\author{Leonardo dos Santos}
\affiliation{Observatoire de l'Universit\'{e} de Gen\`{e}ve, 51 chemin des Maillettes, 1290 Versoix, Switzerland}
\affiliation{Departamento de Astronomia do IAG/USP, Universidade de
S\~{a}o Paulo, Rua do Mat\~{a}o 1226, Cidade Universit\'{a}ria, 05508-900 S\~{a}o Paulo, SP, Brazil}

\author{Stefan Dreizler}
\affiliation{Institut f\"ur Astrophysik, Georg-August-Universit\"at, Friedrich-Hund-Platz 1, 37077 G\"ottingen, Germany}

\author{David Yong}
\affiliation{Research School of Astronomy and Astrophysics, The Australian National University, Cotter Road, Canberra, ACT 2611, Australia}

\author{TalaWanda Monroe}
\affiliation{Space Telescope Science Institute, 3700 San Martin Drive, Baltimore, MD 21218, USA}

\author{Luca Casagrande}
\affiliation{Research School of Astronomy and Astrophysics, The Australian National University, Cotter Road, Canberra, ACT 2611, Australia}

% finish this

\correspondingauthor{Megan Bedell}
\email{E-mail: mbedell@flatironinstitute.org}


\begin{abstract}

% aims & context
The compositions of stars are a critical diagnostic tool for many topics in astronomy such as the evolution of our Galaxy, the formation of planets, and the uniqueness of the Sun. Previous spectroscopic measurements indicate a large intrinsic variation in the elemental abundance patterns of stars with similar overall metal content. However, systematic errors arising from inaccuracies in stellar models are known to be a limiting factor in such studies, and thus it is uncertain to what extent the observed diversity of stellar abundance patterns is real.
% methods
Here we report the abundances of 30 elements with precisions of 2\% for 79 Sun-like stars within 100 parsecs. 
Systematic errors are minimized in this study by focusing on solar twin stars and performing a line-by-line differential analysis using high-resolution, high-signal-to-noise spectra. 
% results
We resolve [X/Fe] abundance trends in galactic chemical evolution at precisions of $10^{-3}$ dex Gyr$^{-1}$ and reveal that stars with similar ages and metallicities have nearly identical abundance patterns. 
Contrary to previous results, we find that the ratios of carbon-to-oxygen and magnesium-to-silicon \added{in solar metallicity stars} are homogeneous to within 10\% throughout the solar neighborhood, implying that exoplanets may exhibit much less compositional diversity than previously thought. 
Finally, we demonstrate that the Sun has a subtle deficiency in refractory material relative to \replaced{$\sim$95\%}{$>$80\%} of solar twins \added{(at 2$\sigma$ confidence)}, suggesting a possible signpost for planetary systems like our own.

\end{abstract}

\section{Introduction}

The photosphere of a Sun-like star acts as a fossil record of the nebular cloud from which the star formed, making spectroscopic measurements of stellar compositions an informative probe of chemical evolution throughout the galaxy. Stellar compositions can also support exoplanet studies: because the star and its planets form side-by-side from the same primordial material, the relative compositions of stars that host different types of planets yield constraints on planet formation processes \citep[e.g.][]{gonzalez97, fischer05, thiabaud15} and may even indicate the detailed physical properties of known planets \citep[e.g.][]{dorn15, santos15, unterborn14, unterborn17a}.

Previous investigations on this topic indicate that stars with similar overall metal content display significant diversity in their elemental abundance patterns \citep{adibekyan12,bensby14,brewer16}. This implies a wide range in the possible properties of exoplanets. In particular, the abundance ratios of C/O and Mg/Si have been found to vary considerably, carrying dramatic consequences for the nature of terrestrial planets formed around these stars \citep{delgado10,petigura11,suarez-andres18}. If, for example, a significant population of stars form from gas with a primordial C/O ratio $>$ 0.8, this could lead to carbon-rich planets, which would carry significant repercussions when interpreting the observed compositions of such planets' atmospheres \citep[see e.g.][]{madhu11}. Similarly, the Mg/Si ratio is a key parameter in models of rocky planet interior structure, with a Mg/Si ratio significantly below 1 or above 2 leading to non-Earth-like balances of compounds like pyroxine and olivine within the planet and potentially altering its geological processes \citep{Sotin2007, carter-bond12}.

Beyond individual elemental abundances or ratios, subtle trends across the full set of measured abundances can also carry information relevant to planets. 
In a differential spectroscopic study of eleven solar twins, \citet[][hereafter M09]{melendez09} found that the Sun has an unusual abundance pattern characteristic of dust condensation. When comparing the Sun to the average abundance pattern of eleven solar twins, M09 observed a deficit of refractory elements relative to volatiles. This trend was quantified as a slope in abundance [X/H] as a function of condensation temperature (\tc), the temperature at which an element is expected to condense under protoplanetary disk conditions \citep{lodders03}. This \tc\ trend has remained a topic of extensive debate, with some studies reproducing M09's findings and others being unable to confirm its existence for various samples of Sun-like stars \citep{ramirez09, ramirez10, GonzalezHernandez2010, GonzalezHernandez2013, Gonzalez2010, Gonzalez2011, Schuler2011b}. 

A variety of explanations for \tc\ trends have been proposed. M09 suggested that the solar \tc\ trend may arise as a result of rocky planet formation. \citet{chambers10} subsequently demonstrated that the Sun's refractory deficit corresponds to 4 Earth masses of terrestrial and chondritic material, confirming the plausibility of this explanation. On the other hand, it is also possible that a positive \tc\ slope could be a signature of late-stage accretion of planetary material \citep[e.g.][]{ramirez11, spina15, oh17}, implying that most solar twins have ingested planets in the past. Additionally, while most studies have focused on potential links between \tc\ trends and planet formation, it is also likely that \tc\ behavior is affected by galactic chemical evolution (GCE) and thereby correlates with stellar age \citep{adibekyan14, nissen15, spina16b}. Alternative causes of the Sun's unusual \tc\ trend include gas-dust segregation in the protoplanetary disk and dust cleansing in the primordial nebula \citep{onehag14, gaidos15}.

It is clear that the detailed abundances of stars, when known at a sufficient level of precision, can strongly inform our understanding of planet formation and the nature of exoplanetary systems. 
However, systematic errors arising from inaccuracies in stellar models, uncertainties in galactic chemical evolution, and small sample sizes are known to be limiting factors in such studies, and thus the robustness of previous results is highly debated \citep{asplund09, fortney12, hinkel16, adibekyan14, nissen15}.

In this work, we focus on solar twin stars to minimize stellar model-driven biases and achieve maximum-precision abundance measurements. These objects are a subset of main-sequence stars with effective temperatures, surface gravities, and overall metallicities very close to the solar values, thus ensuring that the photospheric conditions under which their spectra originate are as similar as possible to the Sun's. As a result, relatively little input from stellar atmospheric models is needed to measure spectroscopic abundances in these stars relative to the solar composition provided that these measurements are done in a strictly differential way. This technique enables us to sidestep the systematic biases introduced by models and achieve abundance measurements with precisions below the level of 0.01 dex, or 2\% \citep{bedell14}.

We present precise abundance measurements for 16 light ($Z \leq 30$) elements in 79 stars. We couple these new measurements with the stellar ages and heavy-element abundances for the same sample presented in \citet{spina17}. The resulting set of abundances for 30 elements across a uniformly analyzed sample of many Sun-like stars enables the investigation at an unprecedentedly detailed level of the behavior of stellar compositions as a function of age, the range of relative concentrations of key planet-forming elements, and the spread of abundance-\tc\ trend behavior.

\section{Data}

Solar twin stars for this study were selected from a variety of sources including color-based dedicated solar twin searches \citep{melendez07,ramirez09,melendez09} as well as twins identified in past surveys \citep{valenti05,baumann10,bensby14}. The general selection criteria used were effective temperature (\teff) within 100 K of solar; surface gravity (\logg) within 0.1 dex of solar; and metallicity (taken as the iron abundance \feh) within 0.1 dex of solar. In some cases our final derived parameters (see Section \ref{ss:ews}) are slightly outside of these bounds.

To achieve sufficient signal-to-noise for high-precision abundance work, we stacked $\ge50$ observations for each star. 
All spectra were taken with the High Accuracy Radial velocity Planet Searcher (\acronym{HARPS}) spectrograph on the 3.6 meter telescope of the European Southern Observatory (\acronym{ESO}), located at La Silla Observatory in Chile \citep{mayor03}.
\acronym{HARPS} is an ultra-stable echelle spectrograph with resolving power R = 115,000 and wavelength coverage between 378 - 691 nm.
A majority of the selected sample were observed by us in the course of a large \acronym{ESO} observing program on \acronym{HARPS} \citep{melendez15}. Other stars had a sufficient number of publicly available spectra in the online \acronym{ESO} Science Archive Facility.\footnote{Data were used from \acronym{ESO} programme IDs 072.C-0488, 074.C-0364, 075.C-0202, 075.C-0332, 076.C-0155, 077.C-0364, 088.C-0323, 089.C-0415, 089.C-0732, 090.C-0421, 091.C-0034, 091.C-0936, 092.C-0721, 093.C-0409, 095.C-0551, 096.C-0499, 097.C-0090, 097.C-0571, 097.C-0948, 183.C-0972, 183.D-0729, 185.D-0056, 188.C-0265, 192.C-0224, 192.C-0852, 289.C-5053, 292.C-5004, and 60.A-9036.}

The \acronym{HARPS} pipeline provides extracted 1-dimensional spectra and radial velocity information automatically. Using these data products, we Doppler-corrected, continuum normalized, and stacked all spectra for a given star to create a single composite spectrum with a very high signal-to-noise ratio (\acronym{SNR}). This procedure was done slightly differently for the blue (378 $< \lambda <$ 530 nm) and red (533 $< \lambda <$ 691 nm) CCD chips. The blue chips were automatically continuum normalized using a 20th-order polynomial fit in IRAF's \texttt{onedspec.continuum} module.\footnote{IRAF is distributed by the National Optical Astronomy Observatory, which is operated by the Association of Universities for Research in Astronomy (AURA) under cooperative agreement with the National Science Foundation.} The red chips, which are more strongly affected by wide bands of telluric absorption lines, were hand-fit in interactive mode. In either case, efforts were made to ensure that all spectra were normalized similarly across stars, since our differential equivalent width technique is more sensitive to star-to-star differences than it is to absolute continuum slopes.

Combined spectra for a total of 79 targets were created in this manner. 
An additional three stars (\hip{19911}, \hip{67620}, and \hip{103983}) were included in the initial solar twin sample, but were later dropped due to spectral contamination by a nearby companion \citep{dosSantos17}. 
The combined spectra of the remaining stars varies in quality depending on the brightness of the target and the number of \acronym{HARPS} spectra observed, but on average a \acronym{SNR} of approximately 800 pix$^{-1}$ at 600 nm is achieved. The minimum and maximum \acronym{SNR} in the sample are 300 and 1800 pix$^{-1}$.

The solar reference spectrum used in this work was created by combining multiple exposures of sunlight reflected from the asteroid Vesta. It was continuum-normalized in the same manner as the target spectra and has \acronym{SNR} $\sim 1300$ pix$^{-1}$ at 600 nm.

Spectra previously obtained with the \acronym{MIKE} spectrograph \citep{bernstein03} and analyzed in \citet{ramirez14} were also used in some parts of this analysis. These spectra have \acronym{SNR} $\sim$ 400 pix$^{-1}$ at 600 nm, resolution R = 83,000 - 65,000 (on blue/red CCDs), and wavelength coverage between 320 and 1000 nm.

\section{Analysis}
\subsection{Stellar parameters and abundances}
\label{ss:ews}

We use a strictly differential line-by-line equivalent width technique, as detailed in \citet{bedell14}, to obtain stellar parameters and abundances. In brief, we measure equivalent widths using IRAF's \texttt{splot} module for each line of interest. We then compute the corresponding abundance for each line using \texttt{MOOG} \added{with stellar models from the Kurucz \acronym{ATLAS9} grid \citep{sneden73, castelli04}}. We subtract the calculated abundances of the solar reference spectrum from each target spectrum before combining lines of a given species to yield a final estimate of the abundance.

As presented in \citet{spina17}, we first measured a set of 98 Fe \I\ and 17 Fe \II\ lines and used these to identify the optimal stellar model parameters (effective temperature \teff, surface gravity \logg, metallicity \feh, and microturbulent velocity \vt). This is done by imposing several criteria on the resulting differential iron abundances of a given star and sampling model parameter space until these criteria are fulfilled. These criteria consist of: (a) slope of the abundance vs.\ reduced equivalent width relation consistent with zero; (b) slope of the abundance vs.\ line excitation potential relation consistent with zero; (c) average abundance of Fe \I\ consistent with that of Fe \II; and (d) derived iron abundance consistent with the input metallicity of the model. The scatter in abundances from the sample of Fe \I\ and Fe \II\ lines are propagated through to define a one-sigma uncertainty range on the derived parameters. The procedure of solving for stellar parameters and their uncertainties was performed automatically using the python package \texttt{$q^2$} \citep{ramirez14}. \deleted{For this analysis, we adopted stellar models from the Kurucz \acronym{ATLAS9} grid \citep{castelli04}. }

\begin{figure}
\centering
\includegraphics[width=\columnwidth]{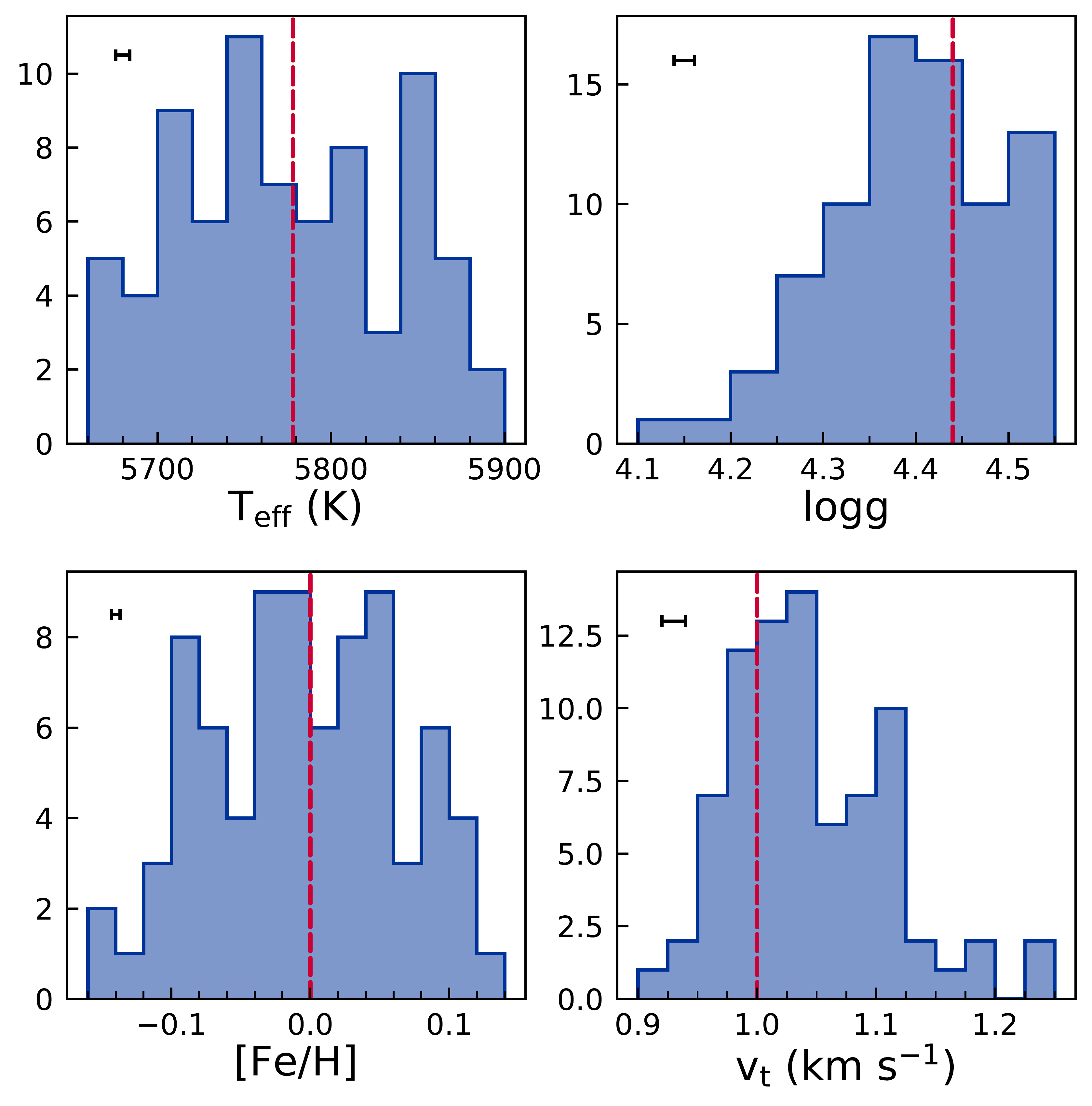}
\caption{Distribution of fundamental stellar parameters (effective temperature \teff, surface gravity \logg, metallicity \feh, and microturbulent velocity \vt) for the 79-star sample. The targets fall within a relatively narrow range around the solar values (vertical dashed lines). The median uncertainty for each quantity is shown as a single 1-$\sigma$ error interval in the upper left corner of the panel.}
\label{fig:params}
\end{figure}

The resulting stellar parameters for the full 79-star sample are shown in Figure \ref{fig:params}. A complete list of the parameters and an analysis of their precision can be found in \citet{spina17}.

Using the same differential equivalent width method, we measured the abundances of 17 additional elements in each spectrum: C, O, Na, Mg, Al, Si, S, Ca, Sc, Ti, V, Cr, Mn, Co, Ni, Cu, and Zn. The line list was adapted from previous works \citep{bedell14, ramirez14}. Hyperfine structure was considered for the V, Mn, Co, and Cu lines using \texttt{MOOG \textit{blends}}. Four elements were measured in multiple species: C \I\ \& CH, Sc \I\ \& Sc \II,  Ti \I\ \& Ti \II, and Cr \I\ \& Cr \II. Oxygen was measured from the \acronym{MIKE} spectra rather than \acronym{HARPS}, as further infrared coverage was needed for the O \I\ triplet. The oxygen abundances were then corrected for departures from local thermal equilibrium \citep{ramirez13}. 

\added{When selecting lines, we preferred relatively strong and well-isolated lines without too many blends, which can be fit with a single Gaussian to determine the equivalent width. For some species, however, such unblended lines are not readily available. In these cases, the equivalent width was measured either by fitting a blend of multiple Gaussians (if the centers of the blended lines are sufficiently separated) or by fitting a single Gaussian while avoiding one contaminated wing. The S \I\ and CH lines used in this analysis were typically composed of multiple closely grouped weak lines of the same species; we found that fitting a single broad Gaussian to the entire set yielded acceptable results.}

Final abundances for each element were obtained by taking an average over the measured abundances for all lines (including all measured species of a given element). Uncertainties on these abundances come from the standard error on the mean, added in quadrature with the propagated effects of the uncertainties on the stellar model parameters. This analysis is automated in the $q^2$ package and can be fully reproduced using the provided list of measured equivalent widths (Table \ref{tbl:ews}). The resulting differential abundances are provided in Table \ref{tbl:abundances}.

In the following analysis, we also use abundance measurements of twelve heavy elements (Sr, Y, Zr, Ba, La, Ce, Pr, Nd, Sm, Eu, Gd, and Dy). These abundances are adopted from \citet{spina17}, which uses the same combined \acronym{HARPS} spectra and the same strictly differential equivalent width method of abundance determination. The resulting dataset of 30 elemental abundances in 79 Sun-like stars is the most extensive set of extremely precise differential abundances produced to date. % too strong??

\begin{figure}
\centering
\includegraphics[width=\columnwidth]{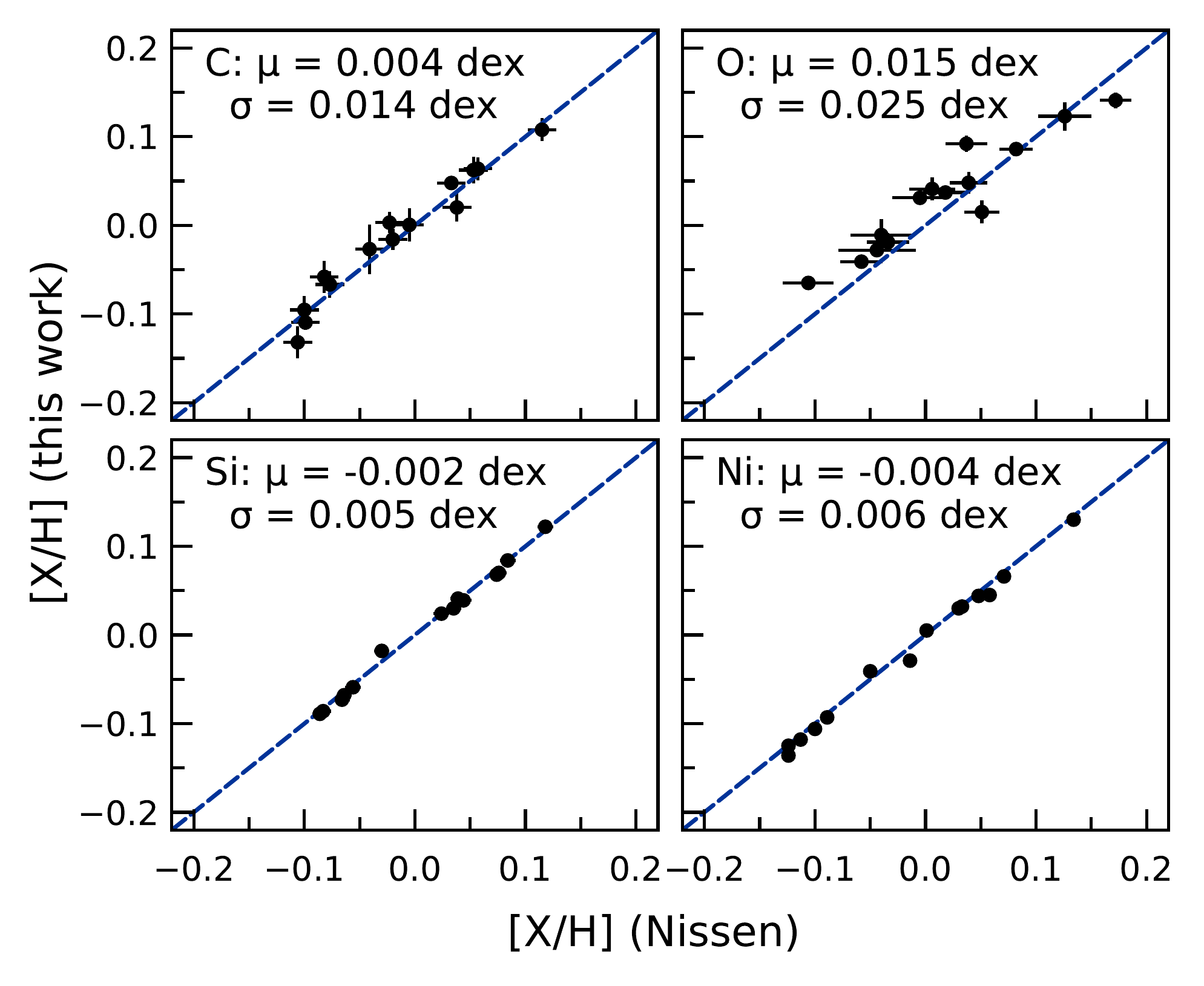}
\caption{Comparison of abundances derived in this work with those from \citet{nissen15} for the 14 overlapping stars in the sample. The mean and standard deviation of the differences are given in each panel. }
\label{fig:nissen}
\end{figure}

The stellar sample considered here overlaps with that of \citet{nissen15} and \citet{nissen16}, who also achieve precise abundances using a similar but independently performed analysis of 21 solar twins. Using the 14 common stars to compare the two data sets, we find extremely good agreement, suggesting that both works do achieve sub-0.01 dex precision in the abundances of most elements (Figure \ref{fig:nissen}). Of the 18 elements measured by \citet{nissen15} or \citet{nissen16} (C, O, Na, Mg, Al, Si, S, Ca, Sc, Ti, Cr, Mn, Fe, Ni, Cu, Zn, Y, and Ba), most display a mean difference and a standard deviation of less than 0.01 dex between the two samples. The elements with the greatest systematic difference between the two samples are O, where this work finds an offset of +0.015 dex relative to \citet{nissen15}, and Ba, where \citet{spina17}'s abundances have an offset of $-0.02$ dex relative to \citet{nissen16}. In the case of oxygen, this discrepancy may be due to the use of different lines and slightly different stellar parameters. For barium, the line lists of the two works are nearly identical, but small differences in the measured equivalent widths and the microturbulent velocity parameters adopted may play a significant role for these strong lines.

\subsection{Stellar ages}

We obtained age estimates for all stars using fits to Yonsei-Yale isochrones \citep{yy}. A joint fit was performed using our measured \teff, \logg, and \feh\ as well as the absolute V magnitude inferred from parallaxes. \textit{Gaia} \acronym{DR1} photometry and parallaxes were used when available \citep{gaiaDR1}; data for the other stars came from the All-Sky Compiled Catalogue and Tycho-2 \citep{ASCC2009, Tycho2000}. We additionally applied metallicity corrections from the [$\alpha$/H] abundances using Mg \I\ as a proxy for $\alpha$. For more detail and a full analysis of the ages obtained, we refer to \citet{spina17}.

\vspace{2mm}

\subsection{Galactic chemical evolution}
\label{s:gce}

To probe the effects of galactic chemical evolution (\acronym{GCE}) on the sample abundances, we investigate the dependence of [X/Fe] on stellar age.

As done in \citet{nissen15} and \citet{spina16b}, we fit trends between abundance and age with a linear model; however, not all stars are well-fit by this model. We chose to exclude 10 stars with ages above 8 Gyr and a visible enhancement in $\alpha$ elements. These apparently belong to a different population, perhaps originating in the thick disk. The excluded stars are: \hip{19911}, \hip{14501}, \hip{28066}, \hip{30476}, \hip{33094}, \hip{65708}, \hip{73241}, \hip{74432}, \hip{108158}, \hip{109821}, and \hip{115577}. 
One additional star, \hip{64150}, was excluded because its abundances of several $s$-process elements are highly anomalous, potentially as a result of past accretion from its close binary companion \citep{dosSantos17, spina17}.

The linear models were fit using an objective function incorporating non-negligible measurement uncertainties in both $(x, y)$ variables \citep{hogg10}. In brief, we minimize the orthogonal distances between the linear model and each data point weighted by the data uncertainties. We assume no covariance in these uncertainties. We additionally incorporate a ``jitter'' or white-noise term added in quadrature with the measurement uncertainties, accounting for the intrinsic star-to-star scatter in abundances. The resulting model has three free parameters for each element: abundance-age slope $m$, intercept $b$, and intrinsic scatter $s$. 

We find best-fit values using the \acronym{BFGS} algorithm as implemented in the \texttt{scipy.optimize.minimize} module, and uncertainties from an \acronym{MCMC} using \texttt{emcee} \citep{emcee} with flat priors on all parameters.

\begin{figure*}
\centering
\includegraphics[trim=14 5 0 10, clip, width=7.5in]{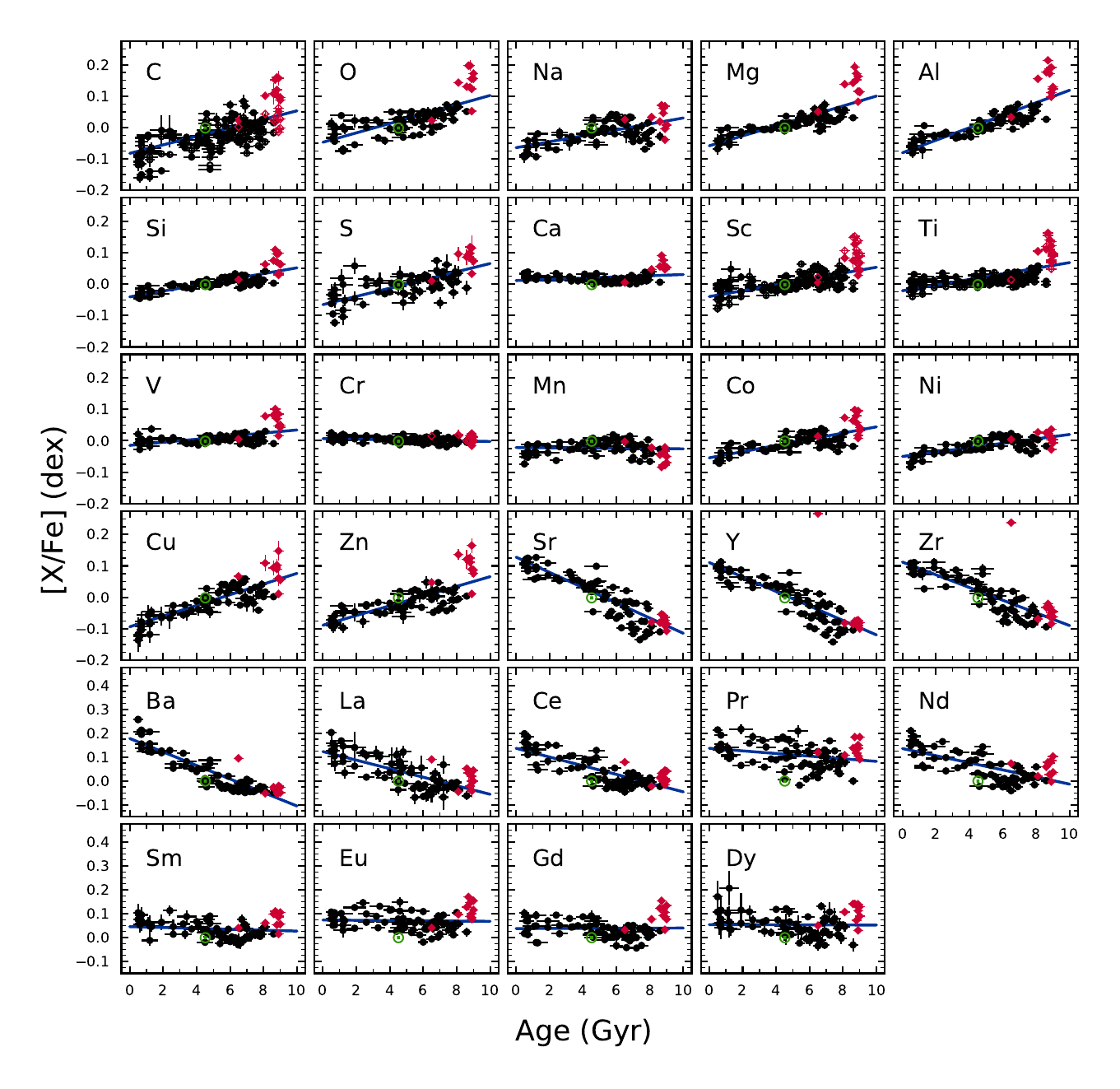}
\caption{Elemental abundances as a function of stellar age. Linear \acronym{GCE} trends (blue) were fit by maximizing a likelihood estimator that accounts for measurement uncertainties in both abundance and age and marginalizes over the unknown level of intrinsic star-to-star variance. Sixty-eight stars (black circles) were included in the linear fits; an additional 11 stars (red diamonds), including a thick-disk-like stellar population at old ages and a star with anomalous abundances potentially due to accretion from its binary companion, were excluded. For elements with more than one species measured, the second species is overplotted with unfilled symbols. The Solar abundances (green) were not considered in the fit, but are nevertheless consistent with it.}
\label{fig:gce}
\end{figure*}

For four elements (C, Ti, Sc, and Cr), abundances were measured using multiple ionization or molecular states. In principle the linear fits should be identical across states. In reality, though, the limited line lists used and systematics-prone nature of spectroscopic analyses mean that different degrees of scatter and even systematic offsets could be present between multiple species of the same element. To account for this possibility, we perform joint fits on the two species with slope $m$ required to be the same and $b, s$ allowed to be different for each one. The resulting best-fit \acronym{GCE} trends are shown in Figure \ref{fig:gce}. Best-fit parameters and uncertainties are given in Table \ref{tbl:gce}.

We note that none of the elements that were measured in multiple species display a systematic offset in their $b$ parameters beyond the level expected from random scatter. The intrinsic scatter parameter $s$ tends to be largest for those species for which few lines were measured, including Na \I\ (4 lines), Cu \I\ (3 lines), O \I\ (3 lines), and the carbon species. This is likely a reflection of the statistical tendency to underestimate uncertainties in the limit of small samples \citep{adibekyan15}. For species with a more extensive list of easy-to-measure lines, like Ti \I\ and Ni \I\ (18 lines each), the intrinsic scatter remains significantly non-zero. This may be a reflection of the true degree of inherent deviations in individual stars from the \acronym{GCE}-predicted average abundance pattern.

\subsection{Condensation temperature trends}
\label{s:tc}

The calculation of trends in elemental abundance with \tc\ is an especially sensitive measurement that can be strongly affected by small inaccuracies in stellar parameters or in the abundances of key volatile elements \citep[e.g.][]{teske16, adibekyan16}. 
Moreover, it is not immediately obvious what model should be used to fit this trend. 
In our analysis, we take a few precautions to maximize the accuracy with which we detect refractory element depletions or enhancements in the sample.

When fitting a linear trend to the abundances, we exclude the volatile elements C, O, S, and Zn. Based on the \tc\ trends seen in solar system meteorites, these elements are not expected to fall along the steepest portion of the refractory-driven \tc\ slope \citep[][and references therein]{chambers10}. \citet{melendez09} accounted for the varying \tc\ behavior of the volatile and refractory elements by fitting with a piecewise linear function, but the location of the line break is not well-determined. Although some subsequent works have included the volatiles in a simple linear fit with the other elements, this approach has its own complications. As noted by \citet{adibekyan16}, elements with under- or over-estimated abundance uncertainties are prone to biasing the overall \tc\ trend if included in the fit. This can especially lead to bias in the case of the most volatile elements (\tc\ $\lesssim 900$ K), as a higher proportion of these elements have uncertain errors due to the small numbers of lines available and the potential systematic effects due to imperfect \acronym{NLTE} corrections. 
In addition, volatiles like C and O may condense under a range of conditions other than their nominal \tc\ values due to their incorporation in a wide variety of compounds, adding a further layer of complexity to the interpretation of these elements from the standpoint of dust condensation processes \citep{lodders03}. For these reasons, we performed our primary \tc\ fits using a single linear model and considering only the 26 elements with \tc\ $>$ 900 K.

It has previously been noted that \acronym{GCE} may affect the \tc\ trends of stars at different ages \citep{adibekyan14}. We make use of our constraints on the abundance--age relations in the sample to subtract off inferred \acronym{GCE} effects from each star's abundances. The \acronym{GCE}-corrected abundances were calculated using the previously derived best-fit \acronym{GCE} trends, with the estimated errors on the \acronym{GCE} slopes propagated to the abundance uncertainties by adding their contributions in quadrature with the existing line-scatter-based uncertainty estimates. The eleven stars that were excluded from the \acronym{GCE} fits were dropped from the post-\acronym{GCE}-correction sample.

We calculated \tc\ trends for every star in the sample by fitting the [X/H] abundances and uncertainties with a non-linear least-squares optimizer as implemented in \texttt{scipy.optimize.least\_squares}. The \tc\ values were taken as 50\% condensation temperatures based on calculations for Solar system composition gas \citep{lodders03}. For each star, a \tc\ slope was calculated both from the ``raw'' [X/H] abundances and also from the abundances after subtracting \acronym{GCE} effects. 

To verify the robustness of our results, we also consider a piecewise linear function as an alternative model. We performed fits in this case including all elements and allowing the break point of the piecewise function to vary as a free parameter. We adopt the slope of the high-\tc\ component as the nominal \tc\ slope for these fits. 

\begin{figure}
\centering
\includegraphics[width=\columnwidth]{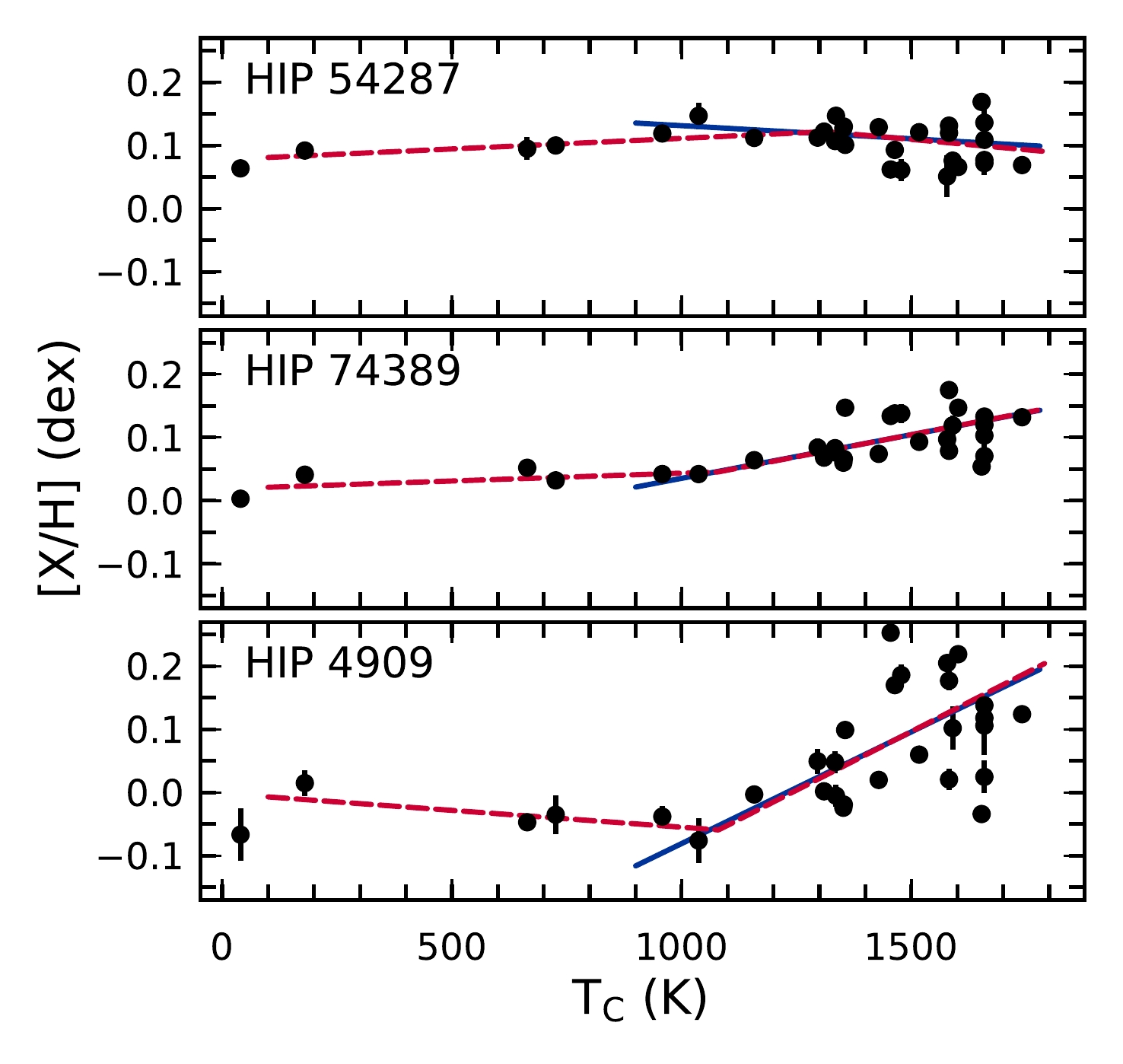}
\caption{Example stars from the sample demonstrating the range of \tc\ trend behavior observed. For each star, differential abundances are shown as a function of the elements' expected condensation temperatures (\tc) in the protoplanetary disk \citep{lodders03}. Best-fit linear (solid blue line) and piecewise linear (dashed red line) fits are also shown.}
\label{fig:tc}
\end{figure}

The resulting \tc\ slopes for each star are presented in Table \ref{tbl:tc}. We plot a few representative stars in Figure \ref{fig:tc} to illustrate the range of behaviors observed.

\begin{figure*}
\centering
\includegraphics[width=\columnwidth]{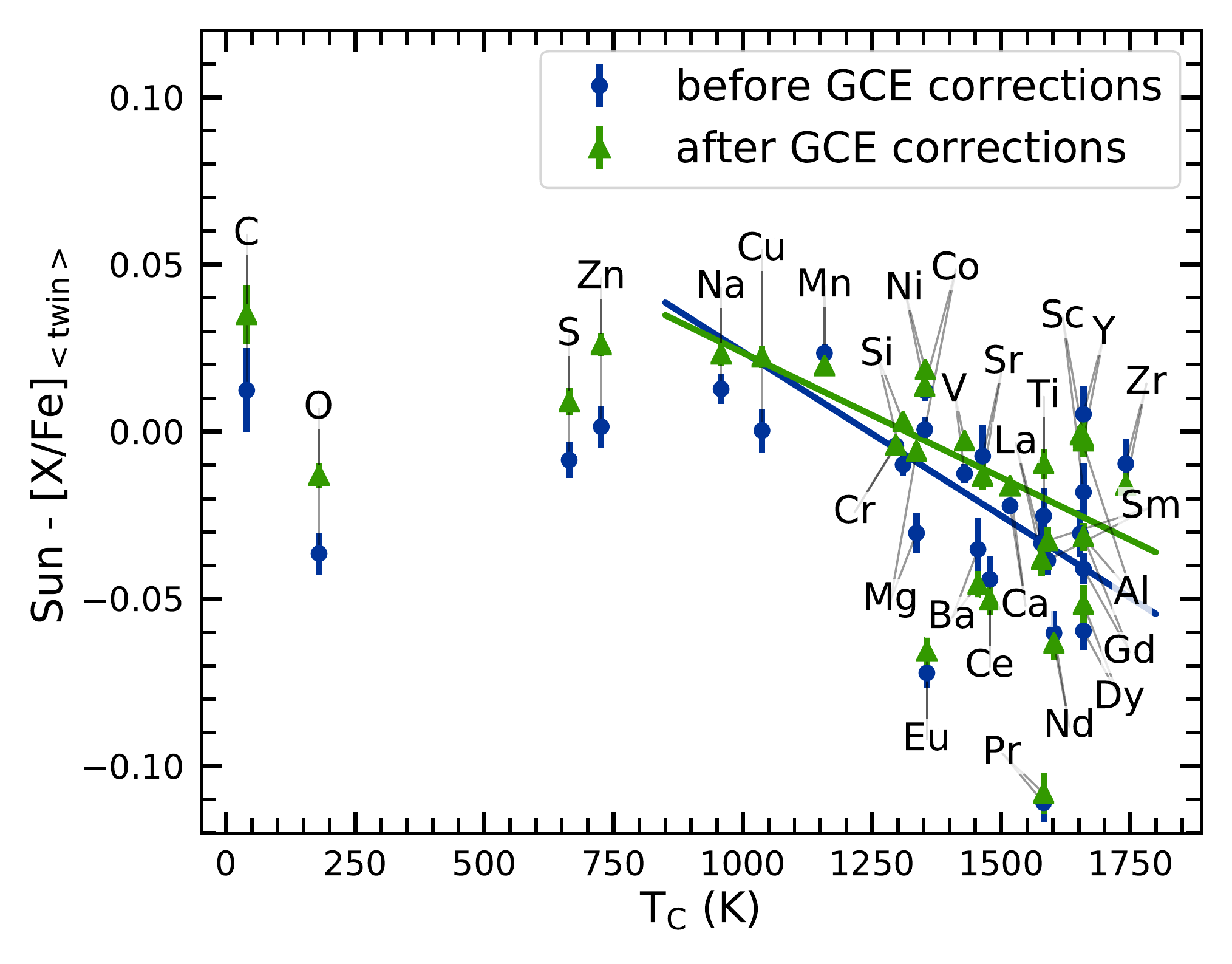}
\caption{The abundance pattern of the Sun compared to the average values in the solar twin sample. Error bars on the abundances are empirically derived as the 1-$\sigma$ error on the mean of the sample. The abundances shown are derived from the full 79-star sample (blue circles) and for the \acronym{GCE}-corrected 68-star sub-sample (green triangles, see text for details). Linear trends are fit to 25 refractory elements and the best-fit lines are shown. Relative to the typical solar twin, the Sun appears deficient in refractory materials (or enhanced in volatiles).}
\label{fig:avgtwin}
\end{figure*}

\subsection{The average solar twin}

Finally, following \citet{melendez09}, we calculate the abundances of a ``typical'' solar twin and compare this abundance pattern to that of the Sun. We calculate the sample average in linear space with the number density of atoms; that is, we take the detailed abundance pattern of the average star to be composed of an equal mixture of materials from the photosphere of every twin. The average abundance ratio of element X to element Y across the full sample of N stars is therefore defined as:

\begin{equation}
 \Big \langle \Big[ \frac{X}{Y} \Big] \Big \rangle = \log_{10} \Big( \frac{1}{N} \sum_{n=0}^{N} 10^{[\frac{X}{Y}]_n} \Big) .
\end{equation}

In Section \ref{s:tc} we used [X/H] rather than [X/Fe] to calculate \tc\ trends, as recommended by \citet{adibekyan16}. For an individual star with well-determined [Fe/H], either abundance ratio will yield virtually the same \tc\ slope with only the offset parameter changing, and by adopting [X/H] we are able to include [Fe/H] as an additional data point. When determining the sample average abundances, though, the propagation of individual stars' [Fe/H] makes a difference to the average calculated abundances (and therefore the resulting \tc\ slope). We made the choice to use [X/Fe] in an attempt to minimize the potential effects of diffusion processes in stars of different ages or slightly different masses. The magnitude of diffusion effects should be much smaller when \added{comparing} an elemental abundance to iron (which diffuses out of the photosphere like the other measured elements) than when comparing to hydrogen (which actually increases in photospheric concentration over time). While some residual effects of diffusion will be present in the ratios of elements with varying masses and ionic charges, it is expected that normalizing for [Fe/H] will reduce these effects to the level of the measurement uncertainties \citep{dotter17}.

In practice, the difference between $\langle [ \frac{X}{Fe} ] \rangle$ and $\langle [ \frac{X}{H} ] \rangle$ is negligible (below 0.005 dex for all elements), and we find the same average \tc\ trend regardless. The trend was fit as in Section \ref{s:tc}, with abundance uncertainties from the scatter among the sample. The resulting \tc\ fits for the sample and for the GCE-corrected sample are shown in Figure \ref{fig:avgtwin}.

\begin{figure*}
\centering
\includegraphics[width=6in]{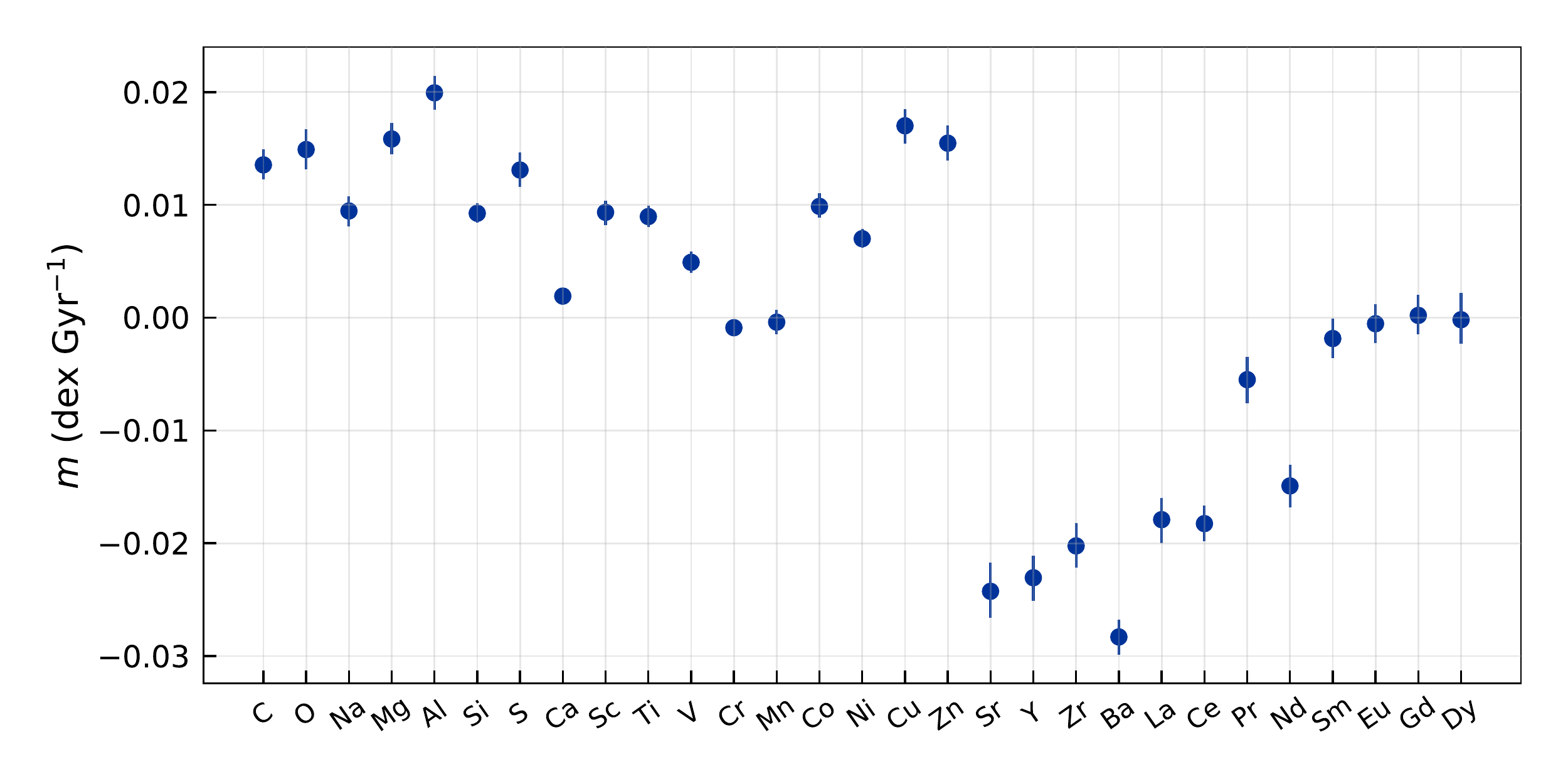}
\caption{\added{[X/Fe] vs. age slopes $m$ for all 29 elements in the sample, ordered by atomic number. The fitting procedure used to derive these parameters is detailed in Section \ref{s:gce}. Noticeable differences are present between heavy ($Z>30$) and light elements, with $s$-process dominated elements exhibiting the most extreme age trends, as previously noted by \citet{spina17}.}}
\label{fig:gce_slopes}
\end{figure*}

\section{Results \& Discussion}
\subsection{Galactic chemical evolution}

The observed \acronym{GCE} trends among our sample (Figure \ref{fig:gce}) generally follow the trends of expected behavior from nucleosynthetic theory. 
The $\alpha$-elements like magnesium and silicon increase with age as the occurrence of type Ia supernovae relative to type II is lower in the earlier universe, with a discontinuity seen in the oldest, thick-disk-like stellar populations \citep{gilmore89}. 
Sodium, as an odd-Z light element, follows a similar trend but shows hints of more complex evolution, as has been previously noted in studies of sodium abundance as a function of metallicity \citep{bensby17}. 
On the other hand, the iron-peak element chromium varies identically to iron through time, as expected. 

\added{Beyond these general categories of nucleosynthetic production, we can resolve more subtle differences among the [X/Fe] vs. age behaviors for individual elements. Figure \ref{fig:gce_slopes} aggregates the observed \acronym{GCE} trends as a function of atomic number. In addition to broad-stroke differences like heavy element ($Z>30$) abundances decreasing with stellar age and lighter elements increasing, the element-to-element scatter in slope exceeds statistical uncertainties even within a nucleosynthetic group. Among the heavy neutron-capture elements, a correlation with atomic number can be seen; this behavior was previously shown in \citet{spina17} and explained in terms of the changing fractional contributions of $s$- and $r$-processes. Among the lighter elements calcium is a notable outlier, having a slope close to zero despite its nominal status as an alpha-element. This behavior, which was also seen by \citet{nissen15}, may be due to its production in sites with longer delay times in addition to type II supernovae. 

A full analysis of the nucleosynthetic information contained in these results would likely require chemodynamical models \citep[e.g.][]{kobayashi11}, but the few examples discussed here serve as a demonstration of the power of these data to constrain nucleosynthetic yields and galactic chemical evolution models.} More broadly speaking, the observed \acronym{GCE} trends place the Sun in context among its neighbors, showing that its chemical composition is characteristic of other solar-metallicity stars formed at the same epoch of galactic history.

\begin{figure}
\centering
\includegraphics[width=\columnwidth]{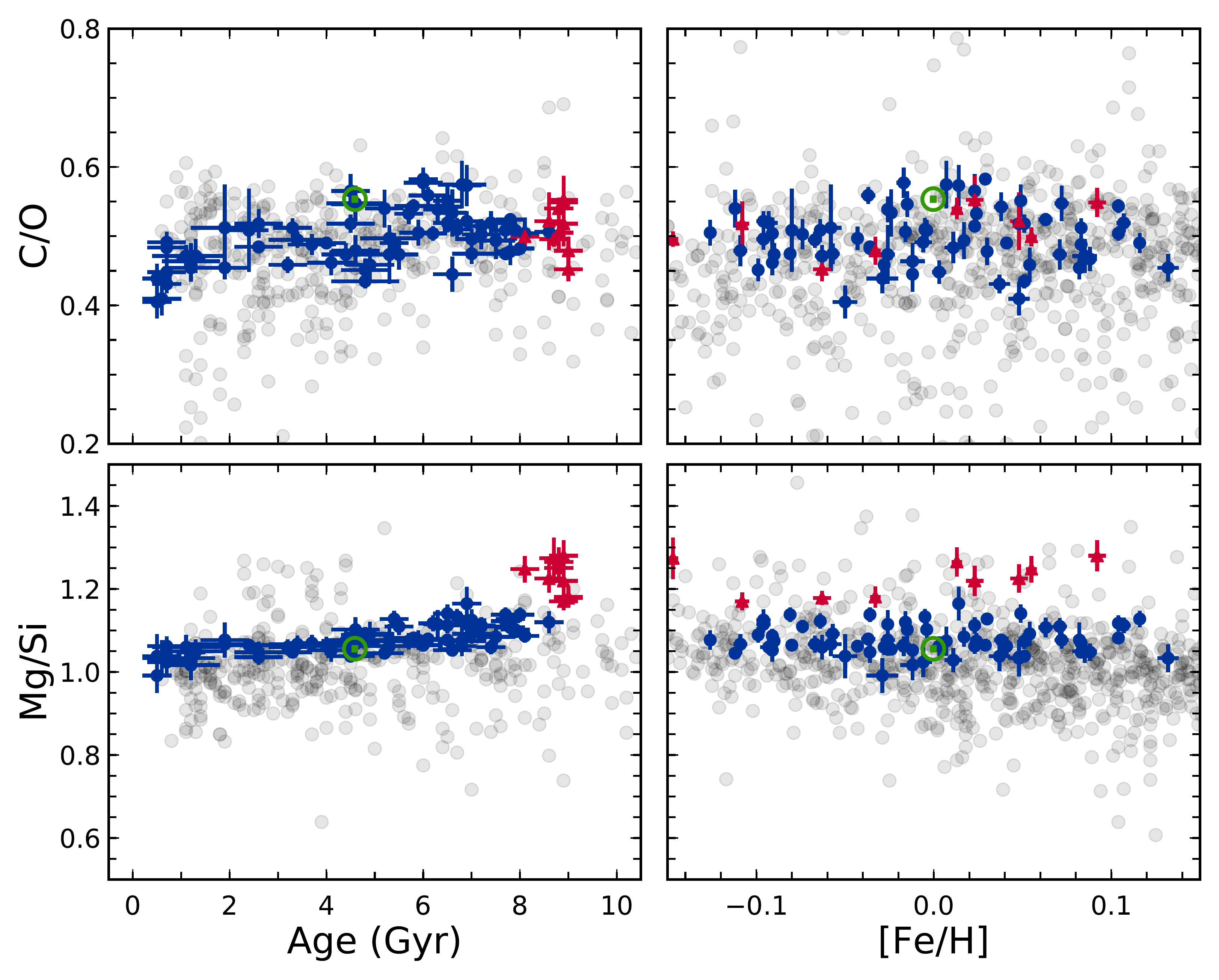}
\caption{Carbon-to-oxygen (C/O, top row) and magnesium-to-silicon (Mg/Si, bottom row) abundance ratios for solar-type stars in the solar neighborhood. Ratios are shown as functions of stellar age (left) and iron abundance (right). Values for a sample of FGK dwarfs in the same metallicity range from \citet{brewer16} are plotted in grey for comparison, with estimated uncertainties of 10\% for C/O and 3.3\% for Mg/Si. By reducing model-dependent bias we find that solar twins (plotted in two populations following the same symbol conventions as Figure \ref{fig:gce}) exhibit lower variance in composition than previously found: in particular, we find no significantly carbon-enriched stars, and all Mg/Si values are consistent with being above unity. We also see a clear evolution in Mg/Si with age, but no trends with bulk metallicity are observed over the range of values investigated.}
\label{fig:ratios}
\end{figure}

\subsection{C/O and Mg/Si ratios}

It is well established that the carbon-to-oxygen (C/O) and magnesium-to-silicon (Mg/Si) abundance ratios play a key role in the formation, atmospheric chemistry, and interior structure in general for all types of planets, and the possibility of plate tectonics and habitability of terrestrial planets in particular \citep{kuchner05, madhu11, oberg11, unterborn14, unterborn17a}.
The broad distribution of C/O and Mg/Si values found in large-scale abundance surveys suggests that a significant fraction of exoplanets are born in environments with very different compositions than that of the primordial solar nebula \citep{delgado10, petigura11, adibekyan12, bensby14, brewer16}, and thus could have strikingly different properties than the solar system planets \citep{bond10, carter-bond12, unterborn17b}. 

We show the derived C/O and Mg/Si abundance ratios from our solar twins abundance study in Figure \ref{fig:ratios}. 
As a comparison sample, FGK stars from the \citet{brewer16} sample with \logg\,$>$ 3.5\,dex and $-0.2 <$ \feh\,$<$\,0.2\,dex are also shown in the figure.
In contrast to previous results, we find very little scatter in these abundance ratios over an 8\,Gyr range in age and for iron abundances within $\pm$0.15\,dex (40\%) of the solar value. 
The C/O values are fully bounded between 0.4 and 0.6, while all the Mg/Si values are consistent with being greater than unity \citep[compare to the solar values of 0.54 and 1.05 respectively][]{asplund09}.
Our results show that these key abundance ratios are quite similar across all Sun-like stars, implying that planetary systems around these stars are born from the same chemical building blocks.

We additionally consider the evolution of these abundance ratios with stellar age and/or metallicity. The C/O ratio appears to evolve very little, as predicted by the GCE simulation of \citet{gaidos15} and in good agreement with past work by \citet{nissen15}. Although the variance of C/O values is smaller than found in \citet{brewer16}, the overall placement of the Sun at the high-C/O end of the distribution agrees well.

On the other hand, the Mg/Si ratio increases with age. This qualitatively agrees with theory: while both Mg and Si are produced by alpha-capture processes in massive stars, there is also non-negligible Si production from Type Ia supernovae \citep[e.g.][]{Burbidge1957}. In keeping with this ongoing Si production, we observe lower Mg/Si in younger stars than in the oldest sample stars. \citet{Adibekyan2015a} found a similar effect when examining the Mg/Si ratio for stars across a wide range in metallicity, with the lowest \feh\ population exhibiting the largest Mg/Si. In this case, we see no significant correlation between \feh\ and Mg/Si, which is not surprising over such a small range in metallicity. Age is apparently a more sensitive tracer of the galactic evolution processes that shape the Mg/Si ratio \added{within this population}.

\added{We can conclude from these results that all nearby planets are formed out of material with Sun-like C/O and Mg/Si ratios if two additional assumptions hold: that the solar twin abundances are representative of stars of different spectral types, and that Mg/Si and C/O ratios do not vary appreciably at non-solar metallicities. Neither of these assumptions is easy to test, as model-based systematics are no longer a negligible contributor to abundances derived for stars of different spectral types and metallicities, or even to a less strictly differential analysis of solar analogs (as in the comparison made in Figure \ref{fig:ratios}). Since the stars in the solar twin sample all have solar metallicity but span a wide range of ages, they must represent a range of nucleosynthetic pathways in galactic star formation, yet all end up with nearly the same Mg/Si and C/O ratios. This suggests that the results found for solar twins can be more widely extrapolated to stars of other metallicities. Other works, however, have found differences in these ratios as a function of metallicity \citep{Adibekyan2015a, Santos2017}. In any case, the vast majority of local stars have approximately solar metallicity \citep[e.g.][]{Casagrande2011}, making these compositional constraints likely to apply for most nearby exoplanet hosts.}

\vspace{2mm}

\subsection{Condensation temperature trends}

\begin{figure}
\centering
\includegraphics[width=\columnwidth]{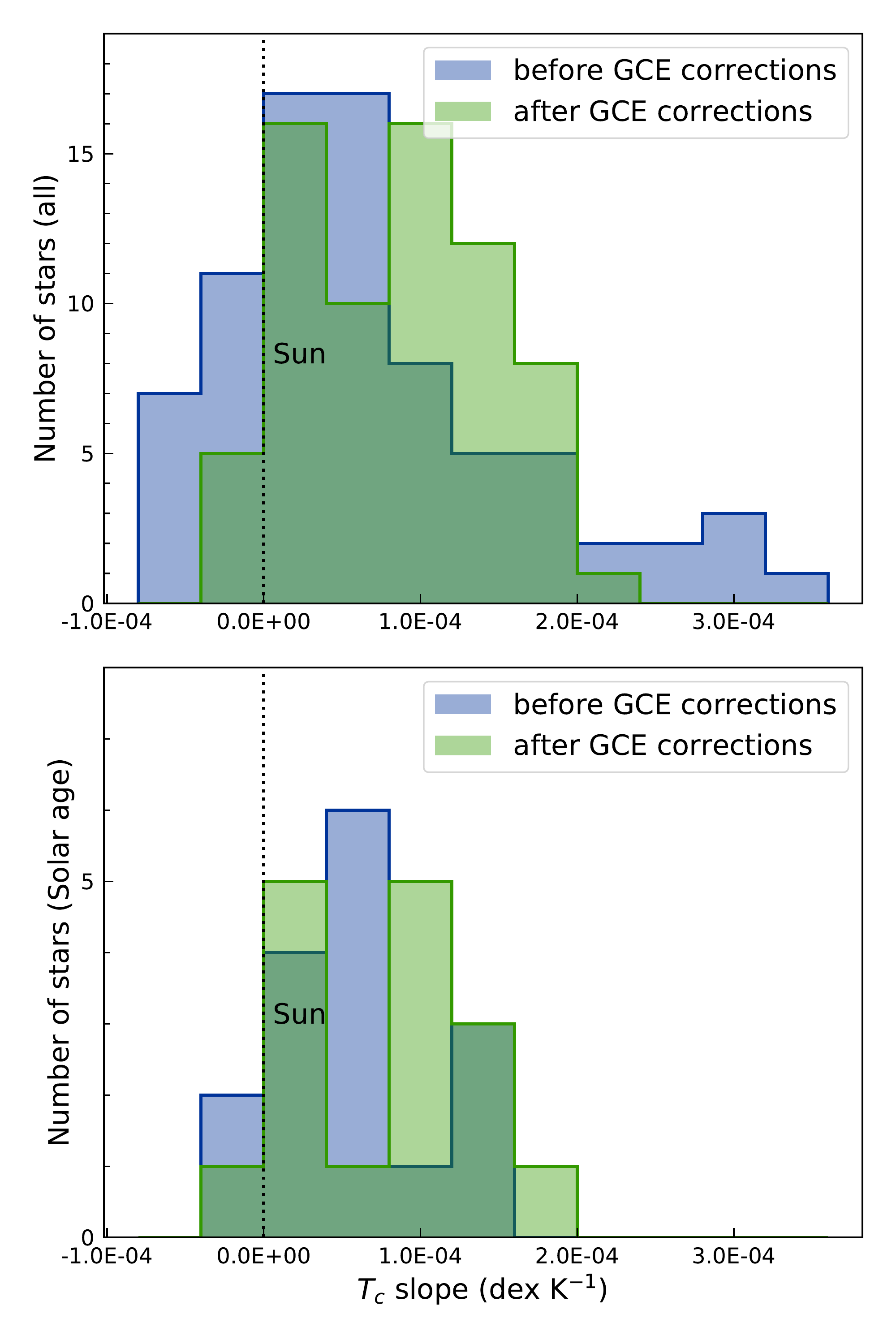}
\caption{Distribution of \tc\ trends in the \replaced{solar twin sample}{full solar twin sample (top) and in the subsample of twins with ages between $3.5-5.5$ Gyr (bottom)}. Slopes in [X/Fe] against \tc\ are shown using abundances from before (blue) and after (green) the \acronym{GCE} corrections shown in Figure \ref{fig:gce}. The solar \tc\ slope, which is zero by definition, lies below \replaced{95\%}{93\%} of the sample (\replaced{87\%}{84\%} at 2$\sigma$ confidence) after correcting for \acronym{GCE}. The Sun's relative deficiency in refractory material may be related to the rarity of the Solar system architecture among exoplanetary systems.}
\label{fig:tchist}
\end{figure}

The results described above suggest that the Sun is entirely typical in its composition. Nevertheless, previous works have debated the existence of one atypical attribute of the solar abundance pattern which may be tied to planet formation: a subtle deficiency in the solar abundances of refractory elements relative to volatiles \citep{melendez09, ramirez09, adibekyan14, nissen15}. 

Despite its important repercussions for the search for Earth-like planets, the existence of an unusual \tc\ trend for the Sun has not been definitively validated due to the difficulty of measuring stellar abundances at the needed level of precision.
The amplitude of the proposed solar abundance trend is much smaller than the sensitivity of most traditional abundance surveys, and it is easily overwhelmed by the systematic errors that plague studies of stars with even moderate differences in their physical parameters.
Abundance trends may also be obscured by the effects of galactic chemical evolution in a sample of stars with different ages \citep{adibekyan14, nissen15, spina16}. 
Our results offer an opportunity to investigate the reality of the proposed solar \tc\ trend for a sample of 79 stars with very precise abundance measurements and well-characterized \acronym{GCE} behavior.

The distribution of \tc\ trends in our sample is shown in Figure \ref{fig:tchist}. Among our \acronym{GCE}-corrected sample of 68 solar twins, we find only \replaced{three}{five} stars with a \tc\ slope at or below the level of the Sun, meaning that the Sun shows a refractory-to-volatile deficiency relative to \replaced{96\%}{93\%} of the sample. 
We estimate the error on this result by resampling the slopes with injected noise based on their measured uncertainties and recording the Sun's position within the resulting simulated distribution. In 95\% of the trials, the Sun lies below at least \replaced{87\%}{84\%} of solar twins. This result does not appear sensitive to the model used for fitting \tc\ trends: in the alternative piecewise linear function, the Sun lies below \replaced{94\%}{82\%} of the \acronym{GCE}-corrected sample.

As shown in the average abundances (Figure \ref{fig:avgtwin}), the neutron-capture elements (including Eu, Pr, Nd, and Ce) tend to be the most discrepant from the \tc\ trend. This may be an indication that our \acronym{GCE} corrections, which are assumed to be simple linear relations with age, do not fully capture the intrinsic variance in abundance patterns. With a more complete marginalization over the abundance contributions of different nucleosynthetic processes, a sharper \tc\ trend with less scatter might emerge. \added{This hypothesis is supported by the lower star-to-star scatter in \tc\ slopes observed in the Solar-age subsample (Figure \ref{fig:tchist}, bottom panel).}

\vspace{10mm}

\replaced{Our ability to interpret these results is limited by current planet detection capabilities. In future works, we hope to interpret the range of \tc\ trends seen in the context of the types of planetary systems around these stars. For the present, we merely note that the rarity of the Sun's abundance pattern and the apparent rarity of Solar system analogs (relative to the frequency of other types of planetary systems), when coupled with the interpretation of \tc\ trends as signatures of dust condensation, make the prospect of Sun-like abundance trends as a marker of Solar-system-like planets worth pursuing in future investigations.}{\subsection{\tc\ Trends and Planets}

All of the stars analyzed in this work have been surveyed for planets with \acronym{HARPS} and in some cases \acronym{HIRES}, making them an excellent sample for investigating possible connections between \tc\ trends and planet formation. To do such an analysis properly, however, requires robust constraints on the occurrence rates of different planet types for the sample. Non-uniform time coverage of the RV observations and star-to-star differences in the manifestations of stellar variability across the sample make such occurrence studies difficult. In particular, terrestrial-mass planets are nearly impossible to detect (or rule out) given the current data. Future work on more robust methods of modeling stellar variability and extracting precise RVs may improve these prospects.

For the present, there are no obvious conclusions to be drawn from the few solar twins which host candidate planets. \hip{11915} and \hip{5301}, which host published Jupiter analogs \citep{bedell15, naef10}, both have \tc\ slopes of around $1.1-1.2\times10^{-4}$ \dexK\ after \acronym{GCE} correction, near the sample average. The same is true of \hip{15527}, which hosts a gas giant planet on an extremely eccentric orbit \citep{Jones2006}. \hip{68468}, which hosts two close-in planet candidates and may have accreted planetary material in the past based on its anomalously high lithium abundance \citep{melendez17}, has a Sun-like \tc\ trend relative to the full solar twin sample. Meanwhile, the radial velocity behavior of the most extreme \tc\ slope stars in the sample do not show any clear anomalies. Given the difficulty of detecting Earth-like planets and the number of confounding factors present in the RVs, it is expected that any connection between terrestrial planet formation and detailed stellar composition should be extremely subtle and will require further analysis.

We leave a more sophisticated analysis of planet properties in this sample to the future. For the present, we merely note that the rarity of the Sun's abundance pattern and the apparent rarity of Solar system analogs (relative to the frequency of other types of planetary systems), when coupled with the interpretation of \tc\ trends as signatures of dust condensation, make the prospect of Sun-like abundance trends as a marker of Solar-system-like planets worthy of further investigation.}

\section{Conclusions}

Using our extremely high-precision measurements of 30 elemental abundances across 79 Sun-like stars in the solar neighborhood, we have constrained galactic chemical evolution trends, investigated the star-to-star variations in key planetary building blocks, and shown that the Sun has an unusual abundance trend relative to the majority of its twins.

The statistics of exoplanetary systems from the \textit{Kepler} mission and radial velocity surveys have demonstrated that the architectures of most other planetary systems are dramatically different than our own, with approximately half of all Sun-like stars hosting likely volatile-rich planets with super-Earth radii in orbits smaller than that of Mercury \citep{winn15, rogers15}. 
The coincidence between the rarity of the Sun's abundance pattern and the solar system architecture offers the intriguing possibility that stellar abundances can be used identify systems harboring planets that are more like the Earth than the typical exoplanet. 
We look forward to more detailed exploration of the connections between stellar abundances and planetary system properties as the sample of bright, spectroscopy-friendly stars with well-characterized characterized planets vastly expands in the upcoming era of \textit{\acronym{TESS}}, \textit{Gaia}, and next-generation radial velocity spectrographs.

\acknowledgements
We thank the many scientists and engineers who made the \acronym{HARPS} observations possible. We also thank Fred Ciesla, Jean-Michel Desert, \added{Bruce Draine,} David W. Hogg, Eliza Kempton, Boris Leistadt, Ben Montet, \added{Melissa Ness,} Poul Nissen, \added{and the anonymous referee} for valuable input.  M.B. acknowledges support for this work from the \acronym{NSF} through the Graduate Research Fellowships Program (grant \#DGE-1144082), the Josephine de Karman Fellowship Trust, the Illinois Space Grant Consortium, and the Lewis and Clark Fund for Exploration and Field Research in Astrobiology. J.L.B.\ acknowledges support for this work from the NSF (grant number AST-1313119), the Alfred P. Sloan Foundation, and the David and Lucile Packard Foundation. J.M.\ and L.S.\ acknowledge the support from \acronym{FAPESP} (2012/24392-2 and 2014/15706-9).

\facilities{ESO:3.6m (HARPS), Magellan:Clay (MIKE)}
\software{\texttt{numpy} \citep{numpy}, \texttt{matplotlib} \citep{matplotlib}, \texttt{IRAF} \citep{iraf}, \texttt{MOOG} \citep{sneden73}, \texttt{q2} \citep{ramirez14}, \texttt{emcee} \citep{emcee}}

\bibliographystyle{apj}
\bibliography{solartwins-aas.bib}%general,myref,inprep}

\newpage

\startlongtable
\begin{deluxetable*}{llcc}
\tablecaption{Contents of equivalent width measurements table.\label{tbl:ews}}
\tabletypesize{\small}
\tablehead{
\colhead{Num} & \colhead{Name} & \colhead{Units} & \colhead{Notes}
}
\startdata
1 & Wavelength & \r{A} & Rest wavelength of line \\
2 & Species &  & Species identifier \\
3 & EP & eV & Excitation potential \\
4 & log($gf$) & dex & Log of the oscillator strength \\
5 & \hip{10175} & m\r{A} & \\ 
6 & \hip{101905} & m\r{A} & \\ 
7 & \hip{102040} & m\r{A} & \\ 
8 & \hip{102152} & m\r{A} & \\ 
9 & \hip{10303} & m\r{A} & \\ 
10 & \hip{104045} & m\r{A} & \\ 
11 & \hip{105184} & m\r{A} & \\ 
12 & \hip{108158} & m\r{A} & \\ 
13 & \hip{108468} & m\r{A} & \\ 
14 & \hip{109821} & m\r{A} & \\ 
15 & \hip{114328} & m\r{A} & \\ 
16 & \hip{114615} & m\r{A} & \\ 
17 & \hip{115577} & m\r{A} & \\ 
18 & \hip{116906} & m\r{A} & \\ 
19 & \hip{117367} & m\r{A} & \\ 
20 & \hip{118115} & m\r{A} & \\ 
21 & \hip{11915} & m\r{A} & \\ 
22 & \hip{14501} & m\r{A} & \\ 
23 & \hip{14614} & m\r{A} & \\ 
24 & \hip{15527} & m\r{A} & \\ 
25 & \hip{18844} & m\r{A} & \\ 
26 & \hip{1954} & m\r{A} & \\ 
27 & \hip{22263} & m\r{A} & \\ 
28 & \hip{25670} & m\r{A} & \\ 
29 & \hip{28066} & m\r{A} & \\ 
30 & \hip{29432} & m\r{A} & \\ 
31 & \hip{29525} & m\r{A} & \\ 
32 & \hip{30037} & m\r{A} & \\ 
33 & \hip{30158} & m\r{A} & \\ 
34 & \hip{30476} & m\r{A} & \\ 
35 & \hip{30502} & m\r{A} & \\ 
36 & \hip{3203} & m\r{A} & \\ 
37 & \hip{33094} & m\r{A} & \\ 
38 & \hip{34511} & m\r{A} & \\ 
39 & \hip{36512} & m\r{A} & \\ 
40 & \hip{36515} & m\r{A} & \\ 
41 & \hip{38072} & m\r{A} & \\ 
42 & \hip{40133} & m\r{A} & \\ 
43 & \hip{41317} & m\r{A} & \\ 
44 & \hip{42333} & m\r{A} & \\ 
45 & \hip{43297} & m\r{A} & \\ 
46 & \hip{44713} & m\r{A} & \\ 
47 & \hip{44935} & m\r{A} & \\ 
48 & \hip{44997} & m\r{A} & \\ 
49 & \hip{4909} & m\r{A} & \\ 
50 & \hip{49756} & m\r{A} & \\ 
51 & \hip{5301} & m\r{A} & \\ 
52 & \hip{54102} & m\r{A} & \\ 
53 & \hip{54287} & m\r{A} & \\ 
54 & \hip{54582} & m\r{A} & \\ 
55 & \hip{62039} & m\r{A} & \\ 
56 & \hip{6407} & m\r{A} & \\ 
57 & \hip{64150} & m\r{A} & \\ 
58 & \hip{64673} & m\r{A} & \\ 
59 & \hip{64713} & m\r{A} & \\ 
60 & \hip{65708} & m\r{A} & \\ 
61 & \hip{68468} & m\r{A} & \\ 
62 & \hip{69645} & m\r{A} & \\ 
63 & \hip{72043} & m\r{A} & \\ 
64 & \hip{73241} & m\r{A} & \\ 
65 & \hip{73815} & m\r{A} & \\ 
66 & \hip{74389} & m\r{A} & \\ 
67 & \hip{74432} & m\r{A} & \\ 
68 & \hip{7585} & m\r{A} & \\ 
69 & \hip{76114} & m\r{A} & \\ 
70 & \hip{77052} & m\r{A} & \\ 
71 & \hip{77883} & m\r{A} & \\ 
72 & \hip{79578} & m\r{A} & \\ 
73 & \hip{79672} & m\r{A} & \\ 
74 & \hip{79715} & m\r{A} & \\ 
75 & \hip{81746} & m\r{A} & \\ 
76 & \hip{83276} & m\r{A} & \\ 
77 & \hip{85042} & m\r{A} & \\ 
78 & \hip{8507} & m\r{A} & \\ 
79 & \hip{87769} & m\r{A} & \\ 
80 & \hip{89650} & m\r{A} & \\ 
81 & \hip{9349} & m\r{A} & \\ 
82 & \hip{95962} & m\r{A} & \\ 
83 & \hip{96160} & m\r{A} & \\ 
84 & sun & m\r{A} & \\ 

\enddata
\tablecomments{Only a portion of this table is shown here to demonstrate its form and content. A machine-readable version of the full table is available.}
\end{deluxetable*}

\pagebreak

\startlongtable
\begin{deluxetable*}{llcc}
\tablecaption{Contents of abundances table.\label{tbl:abundances}}
\tabletypesize{\small}
\tablehead{
\colhead{Num} & \colhead{Name} & \colhead{Units} & \colhead{Notes}
}
\startdata
1 & Star & & Star identifier \\
2 & [C \I /H] & dex & \\ 
3 & u\_[CI /H] & dex & Estimated uncertainty \\ 
4 & [CH /H] & dex & \\ 
5 & u\_[CH /H] & dex & Estimated uncertainty \\ 
6 & [O \I /H] & dex & \\ 
7 & u\_[OI /H] & dex & Estimated uncertainty \\ 
8 & [Na \I /H] & dex & \\ 
9 & u\_[NaI /H] & dex & Estimated uncertainty \\ 
10 & [Mg \I /H] & dex & \\ 
11 & u\_[MgI /H] & dex & Estimated uncertainty \\ 
12 & [Al \I /H] & dex & \\ 
13 & u\_[AlI /H] & dex & Estimated uncertainty \\ 
14 & [Si \I /H] & dex & \\ 
15 & u\_[SiI /H] & dex & Estimated uncertainty \\ 
16 & [S \I /H] & dex & \\ 
17 & u\_[SI /H] & dex & Estimated uncertainty \\ 
18 & [Ca \I /H] & dex & \\ 
19 & u\_[CaI /H] & dex & Estimated uncertainty \\ 
20 & [Sc \I /H] & dex & \\ 
21 & u\_[ScI /H] & dex & Estimated uncertainty \\ 
22 & [Sc \II /H] & dex & \\ 
23 & u\_[ScII /H] & dex & Estimated uncertainty \\ 
24 & [Ti \I /H] & dex & \\ 
25 & u\_[TiI /H] & dex & Estimated uncertainty \\ 
26 & [Ti \II /H] & dex & \\ 
27 & u\_[TiII /H] & dex & Estimated uncertainty \\ 
28 & [V \I /H] & dex & \\ 
29 & u\_[VI /H] & dex & Estimated uncertainty \\ 
30 & [Cr \I /H] & dex & \\ 
31 & u\_[CrI /H] & dex & Estimated uncertainty \\ 
32 & [Cr \II /H] & dex & \\ 
33 & u\_[CrII /H] & dex & Estimated uncertainty \\ 
34 & [Mn \I /H] & dex & \\ 
35 & u\_[MnI /H] & dex & Estimated uncertainty \\ 
36 & [Co \I /H] & dex & \\ 
37 & u\_[CoI /H] & dex & Estimated uncertainty \\ 
38 & [Ni \I /H] & dex & \\ 
39 & u\_[NiI /H] & dex & Estimated uncertainty \\ 
40 & [Cu \I /H] & dex & \\ 
41 & u\_[CuI /H] & dex & Estimated uncertainty \\ 
42 & [Zn \I /H] & dex & \\ 
43 & u\_[ZnI /H] & dex & Estimated uncertainty \\ 

\enddata
\tablecomments{Only a portion of this table is shown here to demonstrate its form and content. A machine-readable version of the full table is available.}
\end{deluxetable*}

\startlongtable
\begin{deluxetable*}{cCCC}
\tablecaption{Best-fit parameters for \acronym{GCE} fits.\label{tbl:gce}}
%\tablewidth{900pt}
%\tabletypesize{\scriptsize}
\tablehead{
\colhead{Species} & \colhead{$m$ (dex Gyr$^{-1}$)} & 
\colhead{$b$ (dex)} & \colhead{$s$ (dex)}
} 
\startdata
C \I & $0.0115 \pm 0.0014$ & $-0.0836 \pm 0.0089$ & $0.0394 \pm 0.0044$ \\
CH & $0.0115 \pm 0.0014$ & $-0.0940 \pm 0.0079$ & $0.0293 \pm 0.0030$ \\
O \I & $0.0088 \pm 0.0014$ & $-0.0260 \pm 0.0075$ & $0.0238 \pm 0.0028$ \\
Na \I & $0.0086 \pm 0.0016$ & $-0.0614 \pm 0.0089$ & $0.0273 \pm 0.0029$ \\
Mg \I & $0.0099 \pm 0.0009$ & $-0.0367 \pm 0.0048$ & $0.0121 \pm 0.0018$ \\
Al \I & $0.0139 \pm 0.0010$ & $-0.0595 \pm 0.0054$ & $0.0156 \pm 0.0019$ \\
Si \I & $0.0063 \pm 0.0006$ & $-0.0308 \pm 0.0033$ & $0.0110 \pm 0.0011$ \\
S \I & $0.0098 \pm 0.0015$ & $-0.0537 \pm 0.0085$ & $0.0240 \pm 0.0035$ \\
Ca \I & $-0.0011 \pm 0.0006$ & $0.0217 \pm 0.0032$ & $0.0089 \pm 0.0011$ \\
Sc \I & $0.0059 \pm 0.0009$ & $-0.0263 \pm 0.0052$ & $0.0159 \pm 0.0022$ \\
Sc \II & $0.0059 \pm 0.0009$ & $-0.0235 \pm 0.0052$ & $0.0205 \pm 0.0025$ \\
Ti \I & $0.0036 \pm 0.0005$ & $-0.0024 \pm 0.0032$ & $0.0119 \pm 0.0014$ \\
Ti \II & $0.0036 \pm 0.0005$ & $-0.0094 \pm 0.0031$ & $0.0115 \pm 0.0014$ \\
V \I & $0.0013 \pm 0.0007$ & $-0.0023 \pm 0.0037$ & $0.0091 \pm 0.0011$ \\
Cr \I & $-0.0016 \pm 0.0003$ & $0.0095 \pm 0.0019$ & $0.0053 \pm 0.0008$ \\
Cr \II & $-0.0016 \pm 0.0003$ & $0.0133 \pm 0.0019$ & $0.0000 \pm 0.0001$ \\
Mn \I & $0.0023 \pm 0.0012$ & $-0.0312 \pm 0.0063$ & $0.0206 \pm 0.0020$ \\
Co \I & $0.0074 \pm 0.0011$ & $-0.0460 \pm 0.0057$ & $0.0178 \pm 0.0020$ \\
Ni \I & $0.0071 \pm 0.0009$ & $-0.0505 \pm 0.0050$ & $0.0172 \pm 0.0018$ \\
Cu \I & $0.0149 \pm 0.0017$ & $-0.0850 \pm 0.0097$ & $0.0244 \pm 0.0030$ \\
Zn \I & $0.0102 \pm 0.0014$ & $-0.0699 \pm 0.0077$ & $0.0224 \pm 0.0029$ \\
Sr \I & $-0.0251 \pm 0.0030$ & $0.1310 \pm 0.0164$ & $0.0574 \pm 0.0060$ \\
Y \II & $-0.0238 \pm 0.0024$ & $0.1135 \pm 0.0130$ & $0.0470 \pm 0.0051$ \\
Zr \II & $-0.0219 \pm 0.0023$ & $0.1179 \pm 0.0125$ & $0.0422 \pm 0.0046$ \\
Ba \II & $-0.0317 \pm 0.0018$ & $0.1897 \pm 0.0093$ & $0.0309 \pm 0.0039$ \\
La \II & $-0.0227 \pm 0.0021$ & $0.1397 \pm 0.0121$ & $0.0350 \pm 0.0046$ \\
Ce \II & $-0.0220 \pm 0.0018$ & $0.1497 \pm 0.0097$ & $0.0305 \pm 0.0039$ \\
Pr \II & $-0.0103 \pm 0.0025$ & $0.1534 \pm 0.0131$ & $0.0451 \pm 0.0050$ \\
Nd \II & $-0.0198 \pm 0.0020$ & $0.1527 \pm 0.0108$ & $0.0360 \pm 0.0038$ \\
Sm \II & $-0.0077 \pm 0.0017$ & $0.0668 \pm 0.0094$ & $0.0226 \pm 0.0032$ \\
Eu \II & $-0.0056 \pm 0.0017$ & $0.0908 \pm 0.0093$ & $0.0300 \pm 0.0033$ \\
Gd \II & $-0.0060 \pm 0.0016$ & $0.0592 \pm 0.0089$ & $0.0279 \pm 0.0028$ \\
Dy \II & $-0.0073 \pm 0.0023$ & $0.0805 \pm 0.0123$ & $0.0332 \pm 0.0043$ \\
\enddata
\end{deluxetable*}

\startlongtable
\begin{deluxetable*}{lCCCC} 
\tablecaption{\tc\ fits for the sample.\label{tbl:tc}}
\tabletypesize{\small}
\tablehead{
\colhead{Star} & \colhead{Slope\tablenotemark{a}} & \colhead{GCE-corrected slope\tablenotemark{a}} &
\colhead{Slope\tablenotemark{b}}  & \colhead{GCE-corrected slope\tablenotemark{b}} \\ 
 & 10^{-4}\ \mathrm{dex\ K}$^{-1}$ & 10^{-4}\ \mathrm{dex\ K}$^{-1}$ & 10^{-4}\ \mathrm{dex\ K}$^{-1}$ & 10^{-4}\ \mathrm{dex\ K}$^{-1}$
 }
\startdata
HIP 10175 & 2.35 & 1.90 & 2.39 & 1.96 \\ 
HIP 101905 & 0.48 & 1.39 & 0.58 & 1.39 \\ 
HIP 102040 & 2.52 & 1.91 & 2.71 & 2.03 \\ 
HIP 102152 & -0.04 & 0.89 & -0.04 & 0.89 \\ 
HIP 10303 & 0.12 & 0.34 & 0.17 & 0.33 \\ 
HIP 104045 & 0.36 & 0.39 & 0.36 & 0.39 \\ 
HIP 105184 & 3.01 & 1.75 & 3.15 & 1.90 \\ 
HIP 108158 & 0.07 & 1.17 & 0.04 & 1.34 \\ 
HIP 108468 & 0.49 & 1.25 & 0.49 & 1.25 \\ 
HIP 109821 & 1.66 & 2.49 & 1.86 & 2.59 \\ 
HIP 114328 & -0.26 & -0.10 & 1.32 & -0.01 \\ 
HIP 114615 & 2.99 & 1.05 & 3.13 & 1.13 \\ 
HIP 115577 & -0.03 & 1.51 & -0.26 & 1.50 \\ 
HIP 116906 & -0.56 & 0.43 & -0.58 & 0.43 \\ 
HIP 117367 & -0.51 & -0.19 & -0.49 & -0.17 \\ 
HIP 118115 & 0.20 & 1.60 & 0.20 & 1.60 \\ 
HIP 11915 & 0.38 & 1.10 & 0.35 & 1.12 \\ 
HIP 14501 & 0.86 & 2.14 & 2.43 & 2.38 \\ 
HIP 14614 & 1.50 & 1.62 & 1.58 & 1.74 \\ 
HIP 15527 & 0.20 & 1.19 & 0.20 & 1.19 \\ 
HIP 18844 & 0.06 & 0.14 & 0.13 & 0.17 \\ 
HIP 1954 & 1.31 & 1.25 & 1.66 & 1.40 \\ 
HIP 22263 & 3.17 & 1.58 & 3.35 & 1.66 \\ 
HIP 25670 & 0.46 & 0.60 & 0.46 & 0.61 \\ 
HIP 28066 & 1.76 & 2.33 & 1.42 & 2.56 \\ 
HIP 29432 & 0.76 & 0.97 & 0.82 & 1.08 \\ 
HIP 29525 & 3.70 & 1.84 & 4.57 & 2.11 \\ 
HIP 30037 & 0.24 & 0.75 & 0.20 & 0.74 \\ 
HIP 30158 & 0.64 & 0.53 & 0.80 & 0.59 \\ 
HIP 30476 & 0.43 & 1.61 & 0.45 & 1.83 \\ 
HIP 30502 & 0.25 & 1.31 & 0.28 & 1.31 \\ 
HIP 3203 & 2.66 & 2.00 & 2.70 & 2.07 \\ 
HIP 33094 & -0.08 & 1.23 & 0.19 & 1.66 \\ 
HIP 34511 & 1.09 & 1.12 & 1.13 & 1.19 \\ 
HIP 36512 & 0.90 & 1.09 & 1.03 & 1.23 \\ 
HIP 36515 & 1.77 & 1.58 & 1.77 & 1.57 \\ 
HIP 38072 & 0.72 & 0.22 & 0.78 & 0.27 \\ 
HIP 40133 & 0.14 & 0.23 & 0.09 & -0.17 \\ 
HIP 41317 & 0.35 & 1.44 & 0.35 & 1.44 \\ 
HIP 42333 & 1.16 & 0.79 & 1.19 & 0.83 \\ 
HIP 43297 & 1.66 & 1.14 & 1.68 & 1.15 \\ 
HIP 44713 & -0.39 & 0.84 & -0.34 & 0.76 \\ 
HIP 44935 & 0.14 & 0.32 & 0.13 & 0.32 \\ 
HIP 44997 & 0.27 & 0.77 & 0.36 & 0.78 \\ 
HIP 4909 & 3.53 & 1.75 & 3.71 & 1.85 \\ 
HIP 49756 & 0.24 & 0.23 & 0.30 & 0.29 \\ 
HIP 5301 & 0.11 & 1.18 & 0.12 & 1.23 \\ 
HIP 54102 & 2.22 & 1.65 & 2.24 & 1.68 \\ 
HIP 54287 & -0.41 & 0.13 & -0.66 & -0.14 \\ 
HIP 54582 & 0.67 & 1.41 & 0.67 & 1.41 \\ 
HIP 62039 & -0.60 & -0.08 & -0.64 & -0.10 \\ 
HIP 6407 & 1.49 & 1.31 & 1.51 & 1.41 \\ 
HIP 64150 & 1.73 & 2.35 & 2.47 & 3.53 \\ 
HIP 64673 & -0.52 & -0.30 & -0.42 & -1.90 \\ 
HIP 64713 & 0.77 & 1.03 & 0.80 & 1.08 \\ 
HIP 65708 & 1.25 & 2.28 & 1.76 & 2.83 \\ 
HIP 68468 & -0.12 & 0.05 & -0.10 & 0.07 \\ 
HIP 69645 & -0.05 & 0.09 & 0.06 & 0.16 \\ 
HIP 72043 & 0.42 & 0.24 & 0.74 & 0.29 \\ 
HIP 73241 & 1.09 & 1.31 & 1.30 & 1.27 \\ 
HIP 73815 & 0.10 & 0.28 & 0.11 & 0.28 \\ 
HIP 74389 & 1.38 & 1.12 & 1.40 & 1.16 \\ 
HIP 74432 & 0.58 & 2.01 & 0.52 & 2.31 \\ 
HIP 7585 & 1.16 & 0.74 & 1.75 & 1.00 \\ 
HIP 76114 & -0.23 & 0.45 & -0.19 & 0.49 \\ 
HIP 77052 & 0.73 & 1.30 & 0.58 & 1.28 \\ 
HIP 77883 & -0.80 & 0.46 & -0.74 & 0.46 \\ 
HIP 79578 & 0.83 & 0.28 & 0.91 & 0.24 \\ 
HIP 79715 & -0.51 & 0.05 & -0.51 & 0.07 \\ 
HIP 79672 & 0.78 & 1.09 & 0.80 & 1.21 \\ 
HIP 81746 & -0.08 & 1.05 & -0.11 & 1.05 \\ 
HIP 83276 & 0.48 & 1.19 & 0.48 & 1.19 \\ 
HIP 85042 & 0.64 & 1.43 & 0.64 & 1.46 \\ 
HIP 8507 & 0.66 & 1.42 & 0.66 & 1.42 \\ 
HIP 87769 & 0.23 & 0.22 & 0.11 & 0.09 \\ 
HIP 89650 & -0.16 & -0.17 & -0.17 & -0.17 \\ 
HIP 9349 & 1.02 & 0.63 & 1.02 & 0.68 \\ 
HIP 95962 & -0.37 & 0.07 & -0.36 & 0.10 \\ 
HIP 96160 & 0.76 & 0.90 & 0.78 & 0.97 \\ 

\enddata
\tablenotetext{a}{from linear fit (see Section \ref{s:tc} for model details)}
\tablenotetext{b}{from piecewise linear model fit}
\end{deluxetable*}

\end{document}